\documentclass[aps,prl,twocolumn,superscriptaddress,preprintnumbers,floatfix]
{revtex4}

\usepackage{graphicx}
\usepackage{epsfig}

\begin{document}

\newcommand{\be}{\begin{eqnarray}}
\newcommand{\ee}{\end{eqnarray}}
\newcommand\del{\partial}
\newcommand\nn{\nonumber}
\newcommand{\Tr}{{\rm Tr}}
\newcommand{\mat}{\left ( \begin{array}{cc}}
\newcommand{\emat}{\end{array} \right )}
\newcommand{\vect}{\left ( \begin{array}{c}}
\newcommand{\evect}{\end{array} \right )}
\newcommand{\tr}{\rm Tr}
\def\conj#1{{{#1}^{*}}}
\newcommand\hatmu{\hat{\mu}}

\title{ Phase of the Fermion Determinant at Nonzero Chemical Potential}

\author{K. Splittorff}
\affiliation{The Niels Bohr Institute, Blegdamsvej 17, DK-2100, Copenhagen {\O}, Denmark}
\author{J.J.M. Verbaarschot$^{1,\,}$}
\affiliation{Department of Physics and Astronomy, SUNY, Stony Brook,
 New York 11794, USA}

\date   {\today}

\date   {\today}
\begin  {abstract}
We show that in the microscopic domain of QCD (also known as the
$\epsilon$-domain) at nonzero chemical 
potential the average phase factor of the fermion determinant is nonzero
for $\mu < m_\pi/2$ and is exponentially suppressed for larger values
of the chemical 
potential. This follows from the chiral Lagrangian that describes
the low-energy limit of the expectation value of the phase factor.
Explicit expressions for the average phase factor are derived using 
a random matrix formulation of the zero momentum limit of this chiral
Lagrangian.

\end{abstract}
\maketitle

{\sl Introduction.} During the past decade, a great deal of progress has been
made in understanding the phase diagram of the QCD partition function
in the chemical potential -- temperature plane. Although early
analytical arguments \cite{pis-wilc} 
clarified the nature of the chiral phase transition
along the temperature axis, 
a detailed quantitative understanding 
could only be achieved by means of lattice QCD simulations (see
\cite{urs} for a review). The situation at nonzero chemical
potential is much less clear. Although perturbative arguments,
model calculations and phenomenological arguments
seem to give a consistent picture \cite{cmqcd}, 
first principle quantitative information is lacking.
The main reason is 
that QCD at nonzero chemical potential  cannot be simulated reliably
in much of the chemical potential -- temperature plane because the
fermion determinant is complex (the sign problem).
Progress has been made around the critical
temperature and small chemical potentials, 
where different lattice QCD approaches seem to converge
\cite{owe,allton,fodor,maria,gupta}. Throughout 
this letter, chemical potential is short for quark chemical potential and
will be denoted by $\mu$.

The question we wish to address in this letter is whether there
exists a parameter domain for which the
phase of the fermion determinant is manageable. Since there is no
sign problem for $\mu =0$, we expect that for sufficiently small
nonzero chemical potential lattice QCD simulations are possible.
The standard argument is that
the number of gauge field configurations required for a converged
calculation diverges exponentially with the volume as $\exp(V \, \Delta F)$,
where $V \Delta F$ is the difference of the 
free energy of typical gauge field configurations
generated by the Monte Carlo algorithm
and the converged free energy. One limit in which the
sign problem remains manageable is when  
$V \Delta F$ remains finite in the
thermodynamic limit. 
We expect that for sufficiently small 
 $\mu$ the free energy difference behaves as
$ \mu^2 V$. For the sign problem to be manageable the chemical
potential then 
has to scale  as $1/\sqrt V$ in the thermodynamic limit.
Such type of limit is well known -- it is the 
microscopic limit or the $\epsilon$-limit of the 
QCD partition function.
In this limit the mass and chemical potential
dependence of the QCD partition function 
is determined  by the zero momentum modes of the Goldstones associated
with the spontaneous breaking of chiral symmetry.

One observable that directly tests the severity of the sign problem
is the expectation value of the phase factor of the fermion determinant.
This phase factor can be expressed as 
the ratio of the fermion determinant and its
complex conjugate
\be
e^{2i\theta} = \frac{{\det}(D +\mu\gamma_0 +m)}
{\det(D^\dagger +\mu\gamma_0 +m)}. 
\ee
Its expectation value is a  
QCD-like partition function with a low energy limit that 
is determined along well-established rules by chiral symmetry and 
gauge invariance \cite{KST,KSTVZ}. In this letter we analyze the 
average phase factor in the microscopic domain of QCD \cite{LS,SV,V,TV}
\be
m_\pi^2 \ll \frac 1{\sqrt V}, \qquad \mu^2 \ll \frac 1{\sqrt V}.
\label{defmicro}
\ee 
In this domain the Compton wave length of the Goldstone modes is much larger
than the linear size of the box, so that
the chiral Lagrangian can be truncated to its zero momentum sector.
In the thermodynamic limit, simple expressions can then be obtained 
using mean field arguments.
At finite volume, the calculations are much more complicated, but
we can exploit the 
equivalence with random matrix theory \cite{SV,V}, 
where recent progress \cite{fyodorov,akemann,O,B,AP,AOSV}
makes it possible
to derive exact results in the microscopic domain. 
Several cases will be discussed: the quenched limit, the
phase quenched limit, and QCD with dynamical flavors. In all cases we 
will find that the average phase factor is nonzero even for large $\mu^2V$ 
provided that $2\mu<m_\pi$. For $2\mu>m_\pi$ the average phase is 
exponentially suppressed with $\mu^2V$. 

Although QCD at nonzero baryon chemical potential has a sign problem,
this is not the case for QCD with two colors, QCD with gauge fields in
the adjoint representation and the phase quenched partition function.
The chemical potential and mass dependence of these partition
functions has been analyzed in great detail by means
of chiral lagrangians or random matrix theory 
\cite{KST,KSTVZ,TV,ss,STV,misha,SplitVerb2,DHSS,SVbos} 
as well as on the lattice \cite{tilo,LatQCDlike,gernot}.
The success
of these calculations suggests that equally impressive lattice 
QCD results can be obtained for the average phase factor. 
We hope that the results presented in this letter 
will encourage such calculations.

The approach introduced in this letter is directly applicable to QCD
at nonzero $\theta$-angle. Fermion sign problems also appear in other
interesting physical systems \cite{CW}. It would be worthwhile to
analyze them along the lines proposed in this letter.

\vspace{1mm}

{\sl General arguments.} 
In this section we will evaluate the 
$\mu$-dependence of the average phase factor 
in the mean field limit. Below we will confirm these results from
the asymptotic limit of the exact expressions for
the microscopic domain. In this domain,
it is natural to work at
fixed topology instead of fixed $\theta$-angle.  
The results presented in this section are for the thermodynamic limit and  
do not depend on the topological charge.

The vacuum energy density 
does not depend on the chemical potential in a phase
that is not sensitive to the boundaries. This is the case in 
the normal phase where 
the chemical potential is below the mass of the lightest
particle with the corresponding charge. For larger $\mu$ there is a
net particle flux in the time direction of the Euclidean torus and the 
free energy depends on the chemical potential \cite{stone}. 
Although in the normal
phase the free energy is $\mu$-independent, the 
excitations of the vacuum are not. 
For a chiral Lagrangian 
the masses of the Goldstone modes for
$\mu < m_\pi/2$ are given by \cite{KSTVZ}
\be
M(\mu) = m_\pi - b \mu,
\ee
where $b$ is the charge of the particles corresponding to  $\mu$. 
In the zero momentum sector, the thermodynamic limit of 
the partition function is therefore given by 
\be
Z = J\prod_k \frac 1{{m_\pi-\mu b_k}} \ e^{-V F},
\label{gen1}
\ee
where the Jacobian, $J$, is from the measure of the Goldstone manifold 
and $F$ is the vacuum energy density, both evaluated at the saddle point.
The prefactor gives a $1/V$ correction to the free energy
density. The prefactor is important if we consider 
the expectation value of the phase factor of the quark determinant which is 
given by the ratio of two partition functions
\be
\langle e^{2i\theta} \rangle_{N_f} = \frac {Z_{N_f+1|1^*}}{Z_{N_f}}.
\label{phase}
\ee
They are defined by ($\langle \cdots \rangle$ refers to quenched averaging) 
\be
Z_{N_f+1|1^*}\! =\! \left \langle 
\frac{{\det}(D +\mu\gamma_0 +m)}
{\det(D^\dagger+\mu\gamma_0 +m)} {\det}^{N_f}(D +\mu\gamma_0 +m)\!
\right\rangle, 
\label{zphase}
\ee 
\be
{\rm and } \qquad Z_{N_f} =\langle {\det}^{N_f} (D+\mu\gamma_0 + m) \rangle.
\label{znf}
\ee
The partition function (\ref{zphase})
contains \footnote{The denominator in (\ref{zphase}) cannot be written as
a convergent bosonic integral. In order to achieve this the denominator and
the numerator have to be multiplied \cite{SVbos}
by $\det(D+\mu\gamma_0 +m)$. However, 
this does not affect the result of the counting argument presented in this
section.}
$N_f+1$ fermionic quarks and  one conjugate
bosonic quark. Assuming 
maximum spontaneous breaking of the axial flavor symmetry,
this results in $N_f+1$ charged fermionic Goldstones composed of
a fermionic quark and a conjugate bosonic anti-quark 
as well as an equal number of  anti-particles with the opposite charge. 
In addition, for
topological charge zero, we have
the usual $(N_f+1)^2$ neutral bosonic
Goldstones and one neutral Goldstone made out
of two bosonic quarks.  
The partition function in the denominator contains $N_f^2$
neutral Goldstones.
In the normal phase, the 
saddle point of the static part of the effective Lagrangian is
$\mu$-independent and neither
the Jacobian nor the free energy do depend on $\mu$.
Using (\ref{gen1}), 
the average phase factor for $\mu<m_\pi/2$ is given by
\be
\langle e^{2i\theta}\rangle_{N_f} =
\frac{(m_\pi^2 -4\mu^2)^{N_f+1}}
{m_\pi^{2N_f+2}}= (1-\frac{4\mu^2}{m_\pi^2})^{N_f+1}.
\label{genres}
\ee
We emphasize that the free energies of $Z_{N_f+1|1^*}$ and $Z_{N_f}$
cancel. Hence for $\mu<m_\pi/2$ the sign problem is not
exponentially hard in the microscopic domain (\ref{defmicro}).
 
This is not the case for $\mu>m_\pi/2$, where the free energy of 
$Z_{N_f+1|1^*}$ is $\mu$-dependent exactly 
as in other theories with charged Goldstone particles \cite{KSTVZ}.
This leads to an exponential suppression of the
average phase factor ($F_\pi$ is the pion decay constant)
\be
\langle e^{2i\theta} \rangle_{N_f} \sim 
e^{-2VF^2_\pi\mu^2(1 -m^2_\pi/4\mu^2)^2}
\label{condensed}
\ee
for $\mu>m_\pi/2$. 
In addition, the Goldstinos with mass
$m_\pi-2\mu$ become exactly massless for $\mu>m_\pi/2$ \cite{KSTVZ} 
so that the leading contributions to the prefactor cancel. We conclude
that the sign problem is not tractable for large $\mu^2F_\pi^2V$ and 
$\mu>m_\pi/2$.

\vspace{1mm}

{\sl Microscopic result.} As we have seen in Eq. (\ref{phase}), 
the expectation value of the phase factor is given by
the ratio of two partition 
functions. We now calculate them in the microscopic limit 
where the scaling variables
\be
\hat{m}=m\Sigma V \quad {\rm and} \quad \hat{\mu}=\mu F_\pi\sqrt{V}  
\ee
are kept fixed for $V\to\infty$. In this limit, the QCD partition function
is equivalent to the large $N$ limit 
of a random matrix theory of $2N\times 2N$ matrices with the
same global symmetries and transformation properties \cite{SV,V}. 
This allows us to
perform the calculations using recent developments
in the method of orthogonal polynomials \cite{O,B,AP,AOSV}.  
Starting from a general expression in \cite{AP}, it can be shown that
the microscopic limit of the  partition functions in (\ref{phase}) 
can be expressed in terms of 
 modified Bessel functions and their Cauchy transforms. For zero
 topological charge we obtain (with $\delta_{\hat m} = \hat m \,d/{d
   \hat m}$)  
\be 
\label{hpZNfmicro}
&&\langle e^{2i\theta}\rangle_{N_f}\sim  \\ 
&&\!\!\frac{1}{Z_{N_f}}\frac{1}{\hat{m}^{N_f(N_f+1)}}\left | \begin{array}{lll}
X^{(0)}(\hat{m};\hat\mu) & \cdots & 
X^{(N_f+1)}(\hat{m};\hat\mu)\\ 
I_0(\hat{m}) & \cdots & \delta_{\hat{m}}^{N_f+1} I_0(\hat{m})\\ 
\vdots & & \vdots \\
\delta_{\hat{m}}^{N_f} I_0(\hat{m}) & \cdots & \delta_{\hat{m}}^{2N_f+1} I_0(\hat{m})\\
\end{array}
\right|, \nn
\ee   
where \cite{LS}
\be
Z_{N_f}\sim \hat m^{-N_f(N_f-1)}\det[\delta_{\hat m}^{k+l}I_0(\hat{m})]_{k,l=0,\ldots,N_f-1}.
\label{zmu0}
\ee

The Cauchy transforms $X^{(k)}(\hat m; \hat \mu)$ are defined by
\be
X^{(k)}(\hat{m};\hat\mu) \equiv  -\frac{e^{-2\hat{\mu}^2}}{4\pi\hat{\mu}^2}
\int {\rm d}^2z \,  
\frac{w(z,z^*,\hat{\mu}) \delta_{z^*}^k I_0(z^*)}{z^2-\hat{m}^2}, 
\label{xk}
\ee
where $w(z,z^*;\mu)$ is the weight function of the random matrix model
in \cite{O}.
The expressions for the Cauchy transform (\ref{xk}) can be rewritten as a
one-dimensional integral following the approach of \cite{SVbos}. 
Next we give explicit results for the
thermodynamic limit of (\ref{hpZNfmicro})  which is obtained from the saddle
point approximation for
$\hat \mu\to \infty  $ and
$\hat m \to \infty $ at fixed $\mu^2/m$.

In the quenched case $(N_f =0)$ a saddle-point approximation of 
(\ref{hpZNfmicro}) gives 
\be
\langle e^{2i\theta}\rangle_{N_f=0} = 
    (1-\frac{4\mu^2}{m_\pi^2})\ e^{0}, \quad 2\mu<m_\pi.
\label{exp-quen-thermo}
\ee  
This result agrees with the mean field arguments given above.
For $\mu > m_\pi/2$ the result is exponentially suppressed 
exactly as in (\ref{condensed}) and with a prefactor that
cancels to leading order in $1/V$.
Notice that the phase factor is only
exponentially suppressed for $\mu>m_\pi/2$.

For $N_f =1$ the thermodynamic limit of the exact microscopic result
(\ref{hpZNfmicro}) is given by 
\be
\langle e^{2i\theta}\rangle_{N_f=1} = 
    (1 -\frac{4 \mu^2}{m_\pi^2})^2  \ e^{0}, \quad 2\mu < m_\pi, 
\label{exp-termo-Nf1}
\ee  
in agreement with the mean field arguments given above.
For $\mu > m_\pi/2 $, at finite volume  the result is  given
by (\ref{condensed}) with a prefactor that cancels to leading order
in $1/V$. 

By now it should be clear that the 
thermodynamic limit of the microscopic 
result (\ref{hpZNfmicro}) reproduces the general formula (\ref{genres}) 
for all values of $N_f$. As further illustration we plot in Fig.~1 
the  average phase factor for $N_f=2$. The dashed curve
represents the result  of Eq. (\ref{hpZNfmicro}) for $mV\Sigma = 4$
and the full curve is its limit for 
$m V \Sigma \to \infty$ at fixed $\mu/m_\pi$.
\begin{figure}[!t]
  \unitlength1.0cm
    \epsfig{file=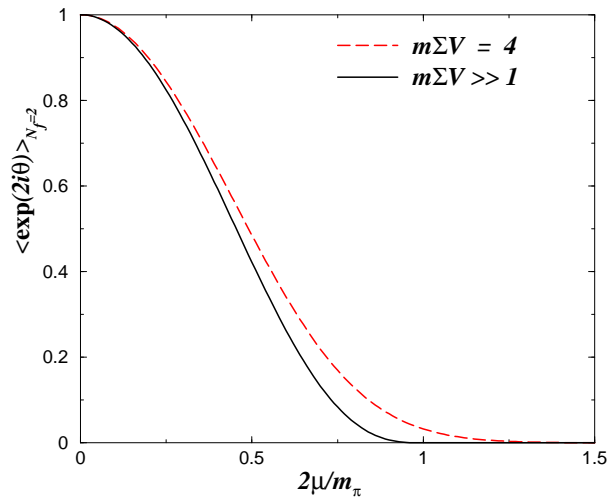,clip=,width=8cm}
  \vspace{-5mm}
  \caption{
  \label{fig:phase} The  average phase factor for 
    two flavor QCD in the
    $\epsilon$-regime as a function of the chemical potential for
    $m\Sigma V=4$ (dashed curve) and $m\Sigma
    V\to\infty$ (full curve).}
\end{figure}
We observe a rapid convergence to the thermodynamic limit especially at
small $\mu$.

Finally, we calculate the average phase factor  
for the phase quenched theory where the phase factor of the dynamical
fermions is ignored. For two flavors it can be expressed as 
\be
\langle e^{2i\theta}\rangle_{1+1^*} = 
\frac{ \langle {\det}^2(D+\mu \gamma_0 +m)\rangle}
{\langle |\det(D+\mu \gamma_0 +m)|^2\rangle} .
\ee
The microscopic limit of both partition functions is well-known
\cite{LS,SplitVerb2}, resulting in the average phase factor
\be
 \langle e^{2i\theta}\rangle_{1+1^*} = \frac{I_0^2(\hat m) - I_1^2(\hat
   m)}{2e^{2\hat\mu^2}\int_0^1 dt t e^{-2\hat{\mu}^2t^2}I_0(\hat{m}t)^2}.
\label{pq}
\ee
In the thermodynamic limit obtained by making a saddle point approximation 
of the integrals in (\ref{pq}) we find
the same result as in the quenched case (\ref{exp-quen-thermo}). 
For $2\mu > m_\pi$ the result is once more given by (\ref{condensed}) but 
with a different prefactor than in the previous two cases.

\vspace{1mm}

{\sl Lattice Simulations.}
Several lattice simulations have studied the fluctuations of the phase 
of the fermion determinant. Both 
$\langle \cos \theta\rangle $  \cite{nakamura} 
and $\langle \theta^2 \rangle $  \cite{allton,ejiri} were
considered. In \cite{allton,ejiri} the
variance of the phase of the fermion determinant was calculated
using the Taylor expansion technique.
To lowest order in an expansion in $ \mu $ the 
average squared phase is given by
 \be
\;\;\langle \theta^2 \rangle -\langle \theta \rangle^2 & =& \left . \mu^2  \del_{\mu_1}\del_{\mu_2} 
\log Z_{1+1^*}
\right|_{\mu_1=\mu_2 =0},
\ee
where
\be
Z_{1+1^*} = 
\langle \det (D+\mu_1\gamma_0 +m) \det( D^\dagger+\mu_2\gamma_0+m)
\rangle.
\ee
In the microscopic limit, this
partition function is given by the the denominator of Eq. (\ref{pq})
with $\mu \to (\mu_1-\mu_2)/2$.
In the thermodynamic limit we obtain ($\mu<m_\pi/2$)
 \be
\langle \theta^2 \rangle-\langle \theta \rangle^2  & =& 2 \frac {\mu^2}{m_\pi^2}+\ldots \ .
\label{vartheta}
\ee
The work of Allton et al. \cite{allton,ejiri} indeed
 suggests that $\langle \theta^2 \rangle \sim \mu^2/m_\pi^2$
but the prefactor appears to be
several times  larger than given by (\ref{vartheta}). There can be several
reason for this discrepancy. 
First, the 
calculations of Allton et al. \cite{allton} were performed close to
the critical temperature 
whereas our results are for zero temperature. In particular, susceptibilities
are expected to be sensitive to the temperature. 
Second, our results have been derived for the $\epsilon$-domain of QCD
whereas the pion mass in \cite{allton} does not satisfy the condition
(\ref{defmicro}).  Third, there could be significant 
ultra-violet contributions to the average squared phase. Although, it can
be shown
along the lines of \cite{STV} that in dimensional regularization the 
$\mu-$dependent terms do not introduce additional ultra-violet 
divergences, for a lattice regularization this is only
the case after the necessary subtractions have been made.
Ultra-violet contributions to the average phase factor 
are expected to behave as $\exp (-V w^2 \Lambda^2)$ with $w$ the width 
of the strip of eigenvalues and $\Lambda$ an
ultra-violet cut-off. Since $w \sim \mu^2$ (see \cite{TV}) ultra-violet
contributions are suppressed in the microscopic limit. 

\vspace{1mm}

{\sl Conclusions.} We have shown that for sufficiently small 
$\mu$ the expectation value of
the phase factor of the
quark determinant can be obtained from chiral perturbation theory. Explicit
expressions have been obtained
in the microscopic  domain where $\mu \sim 1/\sqrt V$ and $m\sim 1/V$.
Our results show a phase transition of $\langle \exp(2i\theta)
\rangle$  at $\mu = m_\pi/2$ with $\langle \exp(2i\theta) \rangle = 0$ 
beyond this point.
In particular, this implies that there is no serious sign problem for
$\mu \sim 1/\sqrt V$ and a physical quark mass. However, 
the sign problem is severe in the {\it chiral limit} for any nonzero
$\mu$ where it is essential for the discontinuity of the chiral
condensate \cite{OSV}. 

Our results have been derived for zero temperature. From the temperature
dependence of the grand potential we expect that  phase fluctuations
initially increase with temperature. 
A deeper understanding of the sign problem could be obtained by
extending the current lattice simulations to lower temperatures 
and quark masses. 
We believe that the confirmation of the analytical results 
for the phase fluctuations  predicted in this letter will be an
important step forward toward a first principles understanding  
of the QCD phase diagram at nonzero chemical potential.

\noindent
{\sl Acknowledgments.} 
We wish to thank G. Akemann, P.H. Damgaard, 
P. de Forcrand, M. L\"uscher, D. Dietrich, J. Osborn
and L. Ravagli 
for valuable discussions. This work was
supported  by U.S. DOE Grant No. DE-FG-88ER40388 (JV) and the 
Carlsberg Foundation (KS).


\begin{thebibliography}{9}


\bibitem{pis-wilc}
  R.~D.~Pisarski and F.~Wilczek,
  Phys.\ Rev.\ D {\bf 29}, 338 (1984).

\bibitem{urs}
U.~M.~Heller, plenary review PoS {\bf LAT2006}. 


\bibitem{cmqcd}
  K.~Rajagopal and F.~Wilczek,
  hep-ph/0011333.


\bibitem{owe}
  P.~de Forcrand and O.~Philipsen,
  Nucl.\ Phys.\ B {\bf 642}, 290 (2002);
  Nucl.\ Phys.\ B {\bf 673}, 170 (2003).

\bibitem{allton}
  C.~R.~Allton {\it et al.},
  Phys.\ Rev.\ D {\bf 66}, 074507 (2002);
  Phys.\ Rev.\ D {\bf 68}, 014507 (2003);
  %
  Phys.\ Rev.\ D {\bf 71}, 054508 (2005).

\bibitem{fodor}
  Z.~Fodor and S.~D.~Katz,
  JHEP {\bf 0203}, 014 (2002);
  JHEP {\bf 0404}, 050 (2004).


\bibitem{maria}
  M.~D'Elia and M.~P.~Lombardo,
  Phys.\ Rev.\ D {\bf 67}, 014505 (2003).

\bibitem{gupta}
  R.~V.~Gavai and S.~Gupta,
  Phys.\ Rev.\ D {\bf 68}, 034506 (2003).


\bibitem{KST}  
J.B. Kogut, M.A. Stephanov, and D. Toublan, 
                      Phys. Lett. B {\bf 464}, 183 (1999).             

\bibitem{KSTVZ}
J.B. Kogut {\it et al.}, 
Nucl.\ Phys.\ B {\bf 582}, 477 (2000).


\bibitem{LS}
  H.~Leutwyler and A.~Smilga,
  Phys.\ Rev.\ D {\bf 46}, 5607 (1992).




\bibitem{SV}           
  E.~V.~Shuryak and J.~J.~M.~Verbaarschot,
  Nucl.\ Phys.\ A {\bf 560}, 306 (1993).

\bibitem{V}                 
  J.~J.~M.~Verbaarschot,
  Phys.\ Rev.\ Lett.\  {\bf 72}, 2531 (1994).

\bibitem{TV}
D.~Toublan and J.~J.~M.~Verbaarschot,
  Int.\ J.\ Mod.\ Phys.\ B {\bf 15}, 1404 (2001).



\bibitem{fyodorov}Y.V. Fyodorov, B. Khoruzhenko and H.J. Sommers,
Ann. Inst. Henri Poincar\'e: Phys. Theor. {\bf 68}, 449 (1998).

\bibitem{akemann}G. Akemann, Phys. Rev. Lett. {\bf 80}, 072002 (2002); J. Phys. A: Math. Gen. {\bf 36}, 3363 (2003).

\bibitem{O}  J.~C.~Osborn,
  Phys.\ Rev.\ Lett.\  {\bf 93}, 222001 (2004).

\bibitem{B} M.C. Berg{\`e}re, hep-th/0311227; hep-th/0404126.

\bibitem{AP} G. Akemann and A. Pottier, 
  J.\ Phys.\ A {\bf 37}, L453 (2004).
 


\bibitem{AOSV}
G.~Akemann, J.~C.~Osborn, K.~Splittorff and J.~J.~M.~Verbaarschot,
 Nucl.\ Phys.\ B {\bf 712}, 287 (2005).

\bibitem{SVbos}
  K.~Splittorff and J.J.M.~Verbaarschot,
  hep-th/0605143.

\bibitem{misha}M. Stephanov, Phys. Rev. Lett. {\bf 76}, 4472 (1996).



\bibitem{SplitVerb2} K.~Splittorff and J.~J.~M.~Verbaarschot,
  Nucl.\ Phys.\ B {\bf 683}, 467 (2004).




 \bibitem{DHSS}
P.~H.~Damgaard, et al. 
  Phys.\ Rev.\ D {\bf 72}, 091501 (2005);
  Phys.\ Rev.\ D {\bf 73}, 074023 (2006);
  hep-th/0604054.




\bibitem{ss}D.~T.~Son and M.~A.~Stephanov,
Phys.\ Rev.\ Lett.\  {\bf 86}, 592 (2001).


\bibitem{STV}
K. Splittorff, D. Toublan, and J.J.M. Verbaarschot,
  Nucl.~Phys. B {\bf 620}, 290 (2002); 
 Nucl. Phys. B {\bf 639}, 524 (2002). 



\bibitem{tilo}
  G.~Akemann and T.~Wettig,
  Phys.\ Rev.\ Lett.\  {\bf 92}, 102002 (2004)
  [Erratum-ibid.\  {\bf 96}, 029902 (2006)];
  J.~C.~Osborn and T.~Wettig,
  PoS {\bf LAT2005}, 200 (2005);
  J.~Bloch and T.~Wettig,
  Phys.\ Rev.\ Lett.\  {\bf 97}, 012003 (2006).
 

\bibitem{LatQCDlike}
  S.~Hands {\it et al.}, 
  Eur.\ Phys.\ J.\ C {\bf 17}, 285 (2000)
  J.~B.~Kogut {\it et al.}, 
  Phys.\ Rev.\ D {\bf 64}, 094505 (2001);
  J.~B.~Kogut, D.~Toublan and D.~K.~Sinclair,
  Nucl.\ Phys.\ B {\bf 642}, 181 (2002);
R.~Aloisio {\it et al.}, 
  Phys.\ Lett.\ B {\bf 493}, 189 (2000);
  S.~Chandrasekharan and F.~J.~Jiang,
  Phys.\ Rev.\ D {\bf 74}, 014506 (2006);
  S.~Muroya, A.~Nakamura and C.~Nonaka,
  Phys.\ Lett.\ B {\bf 551}, 305 (2003).



\bibitem{gernot}
  G.~Akemann and E.~Bittner,
  Phys.\ Rev.\ Lett.\  {\bf 96}, 222002 (2006).



\bibitem{CW}
  S.~Chandrasekharan and U.~J.~Wiese,
  Phys.\ Rev.\ Lett.\  {\bf 83}, 3116 (1999).





\bibitem{stone}
  J.~B.~Kogut, {\it et al.}, 
  Nucl.\ Phys.\ B {\bf 225}, 93 (1983).



 


\bibitem{nakamura}
  Y.~Sasai, A.~Nakamura and T.~Takaishi,
  Nucl.\ Phys.\ Proc.\ Suppl.\  {\bf 129}, 539 (2004).
\bibitem{ejiri}
  S.~Ejiri,
  Phys.\ Rev.\ D {\bf 69}, 094506 (2004);
  Phys.\ Rev.\ D {\bf 73}, 054502 (2006).

\bibitem{OSV}
  J.~C.~Osborn, K.~Splittorff and J.~J.~M.~Verbaarschot,
 Phys.\ Rev.\ Lett.\  {\bf 94}, 202001 (2005).




\end{thebibliography}
\end{document}